\documentstyle[array,multicol,prb,aps,psfig]{revtex}
\draft
\tighten
\begin{document}
%\vss \special{"90 rotate 4 72 mul .55 -72 mul moveto /Times-Roman findfont
%20 scalefont setfont 0.3 setgray (NOT FOR DISTRIBUTION!) show grestore}

\title{Vortex Density of States and Absorption in Clean Layered Superconductors.}
\author{A.~ A.~ Koulakov$^1$ and A.~I.~Larkin$^{1,2}$}
\address{{}$^1$Theoretical Physics Institute, University of Minnesota,
Minneapolis, Minnesota 55455 \\
{}$^2$Landau Institute for Theoretical Physics, 117334 Moscow, Russia}

\date{\today}

\maketitle
\begin{abstract}
We study the spectrum of the states localized in the
vortex cores in the mixed state of clean layered superconductors. 
S-wave coupling is assumed. It is found that in a large region of parameters 
adjacent to the superclean case a new universal (i.e. independent of
the density of impurities) class of level statistics arises.
It is the circular unitary random matrix ensemble.
The density of states for such conditions is calculated.
The absorption resulting from the Landau-Zener transitions
between these levels is different from the classical result
for an isotropic three-dimensional system.
\end{abstract}

%\narrowtext
\begin{multicols}{2}

Vortex cores in superconductors provide the low energy states within the superconducting gap. 
These states are therefore important for the absorption of 
low frequency electromagnetic field.\cite{KK,Feigelman97,Larkin97,Koulakov98}
Of particular interest are such states in layered s-wave superconductors, 
\cite{Feigelman97,Guinea95,Kravtsov98} that mimic
one of the features of high-$T_c$ superconductors (HTSC) - their large anisotropy.
As tunneling between layers is slow, such systems can be considered
quasi two-dimensional (2D). 

% A possible example of such systems may be the organic superconductors. 

%---------------------------------------------------------------------------------
%
%       Fig.05
%
%---------------------------------------------------------------------------------
\begin{figure}
%\vspace{-0.2in}
\centerline{
\psfig{file=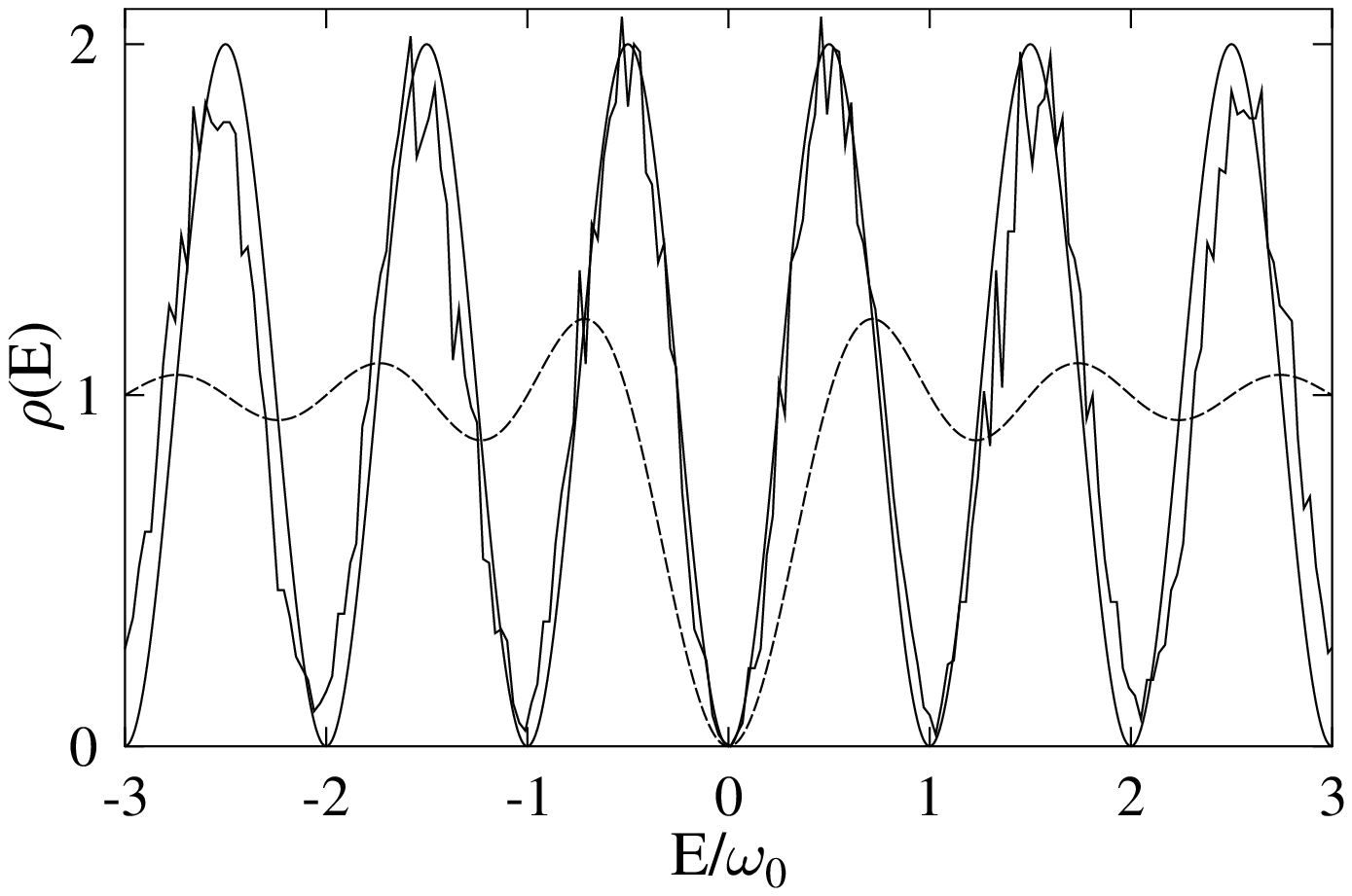,height=1.7in,bbllx=83pt,bblly=105pt,bburx=476pt,bbury=365pt}
}
\vspace{0.1in} 
\setlength{\columnwidth}{3.2in}
\centerline{\caption{DOS near the Fermi level in clean system.
The result for class C of the the 
random-matrix theory is shown by the dashed line.
~\protect{\cite{Kravtsov98,Altland}} 
Eq.~(\protect{\ref{IntroDOS}}) is shown by the solid line. 
The figure also displays the result of the numerical solution
of the system (\protect{\ref{MatrixEq}}) (see below)
averaged over $3000$ realizations of disorder (\ref{Disorder}), 
for $\theta=1$, $n_i\xi^2=4$ (four impurities per vortex core),
$\xi/\lambda_{\rm F}=50$. In the calculation we have
used $100$ basis CDM states.
\label{fig05}
}}
\vspace{-0.1in}
\end{figure}

In the absence of disorder the spectrum of the Caroli - de Gennes - Matricon (CDM) 
states localized in a 2D vortex is equidistant: $E_n^0 = -\omega_0(n - 1/2)$, 
$\omega_0 \sim 1 / m\xi^2 \sim \Delta^2/E_{\rm F}$,
where $n$ is an integer; $\xi$, $m$, $\Delta$, $E_{\rm F}$ 
are the superconducting correlation length, mass of electron,
superconducting gap, and Fermi energy
respectively. Here and below we assume $\hbar=1$, and measure energies from the Fermi level.
In the presence of a small number of the short-range impurities the spectrum retains some of its periodicity. 
In particular in the {\em superclean} regime ($1/\tau \ll \omega_0$) 
such impurities displace the levels with even and odd $n$ in the opposite directions 
by an equal amount.\cite{Larkin97} 
This implies that the spectrum is $2\omega_0$ periodic.
This result holds if the impurities are strong, i.e. their Born parameter $\theta \gg \left( \Delta/E_{\rm F}\right)^{1/2}$.

The levels are shifted only if an impurity is present inside the vortex core. 
Because in the superclean regime the density of impurities is small 
such vortices are rare. Hence, the shape of DOS in the superclean regime 
depends on the density of impurities and is not universal. 

As the amount of disorder is increased the {\em clean} regime ($\omega_0 \ll 1/\tau \ll \Delta$) is approached. 
In this regime the vortex cores can be viewed simplistically as small granules of disordered normal metal.~\cite{Bardeen65}
Therefore one can expect the energy level statistics of Wigner-Dyson type.\cite{Mehta91,Efetov98} 
More precisely, due to the superconductor surrounding the ``granule'',
one can expect level statistics of class C in the nomenclature of Ref ~\onlinecite{Altland}.
This is indeed the case as shown in Ref.~\onlinecite{Kravtsov98}, 
for layered superconductors with a white noise disorder potential, i.e. for vanishingly small
Born parameter $\theta$.
The corresponding density of states (DOS) is shown in Fig.~\ref{fig05}
by a dashed line. The suppression of DOS on the Fermi level is due to the particle-hole 
symmetry ($E \rightarrow -E$) pertinent to the quasiparticles.

In this work we study the clean superconductors with strong short-range impurities:
\begin{equation}
V_{\rm imp}(\bbox{r}) = \sum_i V_{i} 
\delta \left( \bbox{r} - \bbox{a}_i \right),
\label{Disorder}
\end{equation}

\vspace{-0.1in}
\noindent
where $V_i$ and $\bbox{a}_i$ are the strength and the position of the
$i$-th impurity respectively, 
while $\delta(\bbox{r})$ is the function with spatial extent comparable to $\lambda_{\rm F}$.
We assume that the impurity scattering phase $\theta=m(\overline{V_i^2})^{1/2}$ is of the order of one.
We address the question of whether the $2\omega_0$ periodicity of the spectrum, specific for the superclean regime,
persists in the clean regime. We show that in the large subregion of the latter defined by
$\theta^2\Delta\sqrt{\Delta/E_{\rm F}} \gg 1/\tau \gg \omega_0$, the spectrum is indeed $2\omega_0$ periodic.
Thus it is effectively defined by positions of the two nearest levels, with all the others multiplied periodically up and down. 
The DOS in our case is given by
\begin{equation}
\rho(E)=\frac{2}{\omega_0} \sin^2\left(\frac{\pi E}{\omega_0}\right).
\label{IntroDOS}
\end{equation}  
It is shown in Fig.~\ref{fig05} by solid line.
Due to the particle-hole symmetry of the spectrum and the periodicity, the average DOS also defines the 
level spacing distribution $p(s)$ through the relation $p(s) = \rho(s/2)/2$, where $s=2E$ is the level spacing. 
The distribution (\ref{IntroDOS}) is then consistent with the unitary circular random 
matrix ensemble~\cite{Mehta91} for the two independent levels.
This is in contrast to the findings of Ref.~\onlinecite{Kravtsov98} where white noise disorder is considered
($\theta \rightarrow 0$). 
We conclude therefore that in the above limit a new universal (i. e. independent of the density of impurities)
class of level statistics arises.

We calculate also the dissipative component
of conductivity due to the Landau-Zener transitions:\cite{Feigelman97,Larkin97,Koulakov98}
\begin{equation}
%\eta = \hbar \frac{k_{\rm F}^2}{\omega_0\tau} \ln\frac{1}{\theta} 
%\label{Viscosity}
\sigma_{xx} = 32 \sigma_nH_{c2}\ln \theta^{-1} /B(\omega_0\tau)^2.
\label{Viscosity}
\end{equation}
Because in the clean case the parameter $\omega_0 \tau \ll 1$
this result obtained for the layered superconductors
is different from the isotropic three-dimensional case:~\cite{KK,Bardeen65}
\begin{equation}
\sigma_{xx} \approx \sigma_nH_{c2}/B.
\label{BardeenStephen}
\end{equation}

The two-component wavefunctions of excitations $\hat{\psi}$ ($\psi_1=u$, $\psi_2=v$)  
satisfy Bogolubov - de Gennes equations:\cite{DeGennes89}
\begin{equation}
\left[ \sigma_z \left( H_0 +V_{\rm imp}\right) + \sigma_x {\rm Re}\Delta(r) + 
\sigma_y {\rm Im} \Delta(r) \right] \hat{\psi} = E \hat{\psi}.
\label{Bogolubov}
\end{equation}
Here $H_0=p^2/2m-E_{\rm F}$, $\left\{\sigma_x\right.$, $\sigma_y$, %
$\left.\sigma_z\right\}$ are the Pauli matrices, and $E$ is the 
excitation energy.  
In the absence of impurities the wavefunction of the $n$-th CDM state is given by 
\begin{equation}
\hat{\psi}_n=
\frac{e^{-K\left( r\right)}}{2\sqrt{\xi\lambda_{\rm F}}}
\left(
\begin{array}{l}
{\displaystyle
e^{in\phi}J_n\left( k_{\rm F} r\right)
} \\ 
{\displaystyle
-e^{i\left(n-1\right)\phi}J_{n-1}\left( k_{\rm F} r\right)
}
\end{array}
\right).
\label{WaveFunctions}
\end{equation}
Here $J_n\left( x\right)$ is the Bessel function
and $K(r) = \Delta r/v_{\rm F} = r/\xi \pi$. 
Here and below we assume that the 
absolute value of the order parameter is constant 
within the vortex core at small temperatures, 
adopting the Kramer-Pesch ansatz.~\cite{Kramers74}
If impurities are present the excitation wavefunction 
can be found as a mixture of different
CDM states with coefficients $C_m$:
\begin{equation}
\hat{\psi} = \sum_m C_m \hat{\psi}_m.
\end{equation}
$C_m$ satisfy the general perturbation 
theory equation
\begin{equation}
(E^0_n-E)C_n + \sum_{i,m} A^i_{nm} C_m = 0,
\label{MatrixEq}
\end{equation} 
where $A^i_{nm}$ is the matrix element produced by one 
short-range impurity between two 
CDM states $n$ and $m$. The expression for this matrix element 
follows from Eqs.~(\ref{Disorder}), (\ref{Bogolubov}), 
and (\ref{WaveFunctions}):
\begin{equation}
\begin{array}{l}
{\displaystyle A^i_{nm} = V_ie^{-2K(a_i)+i\phi_i(m-n)} 
\left[ J_n\left( k_{\rm F} a_i\right) J_m\left( k_{\rm F} a_i\right)
\right. } \\
{\displaystyle 
\left. -J_{n-1}\left( k_{\rm F} a_i\right) J_{m-1}
\left( k_{\rm F} a_i\right) \right] /{\lambda_{\rm F}\xi}. }
\end{array}
\end{equation}
Here $(a_i, \phi_i)$ is the position of the $i$-th impurity
in the polar coordinates. 
In Eq.~(\ref{MatrixEq}) we ignore the influence of impurities on the order parameter.
The summation index $m$ in 
Eq.~(\ref{MatrixEq}) runs from $-N$ to $N \approx \Delta/\omega_0$,
where $N$ is a very large number. We will assume further $N$ to be infinite and perform a Fourier transform from
integer index $m$ to the real variable $\phi$, using 
$
\Psi(\phi) = \sum_m C_m e^{-i\phi m}.
$
After this transformation Eq.~(\ref{MatrixEq}) acquires the following 
form
\begin{equation}
-i\omega_0\frac{\partial \Psi}{\partial \phi} + 
\int_0^{2\pi} L(\phi, \phi') \Psi(\phi') d\phi'
=\left(E-\frac{\omega_0}{2}\right)\Psi(\phi), 
\label{ExactIntegralEq}
\end{equation}
where the kernel $L(\phi, \phi')$ describes scattering by impurities:
\begin{equation}
\begin{array}{ll}
{\displaystyle L(\phi, \phi')} & 
{\displaystyle 
= \sum_i \frac {V_ie^{-2K(a_i)}}{2\pi\lambda_{\rm F} \xi}
\left[ 1- \exp\left( i\phi'-i\phi\right)\right]
} \\
{} &
{\displaystyle
\exp \left(-ik_{\rm F}a_i\left[ \sin(\phi+\phi_i) - \sin(\phi'+\phi_i) \right]
\right)}
\label{ScatteringMatrix}
\end{array}
\end{equation}

%---------------------------------------------------------------------------------
%
%       Fig.10
%
%---------------------------------------------------------------------------------
\begin{figure}
\centerline{
\psfig{file=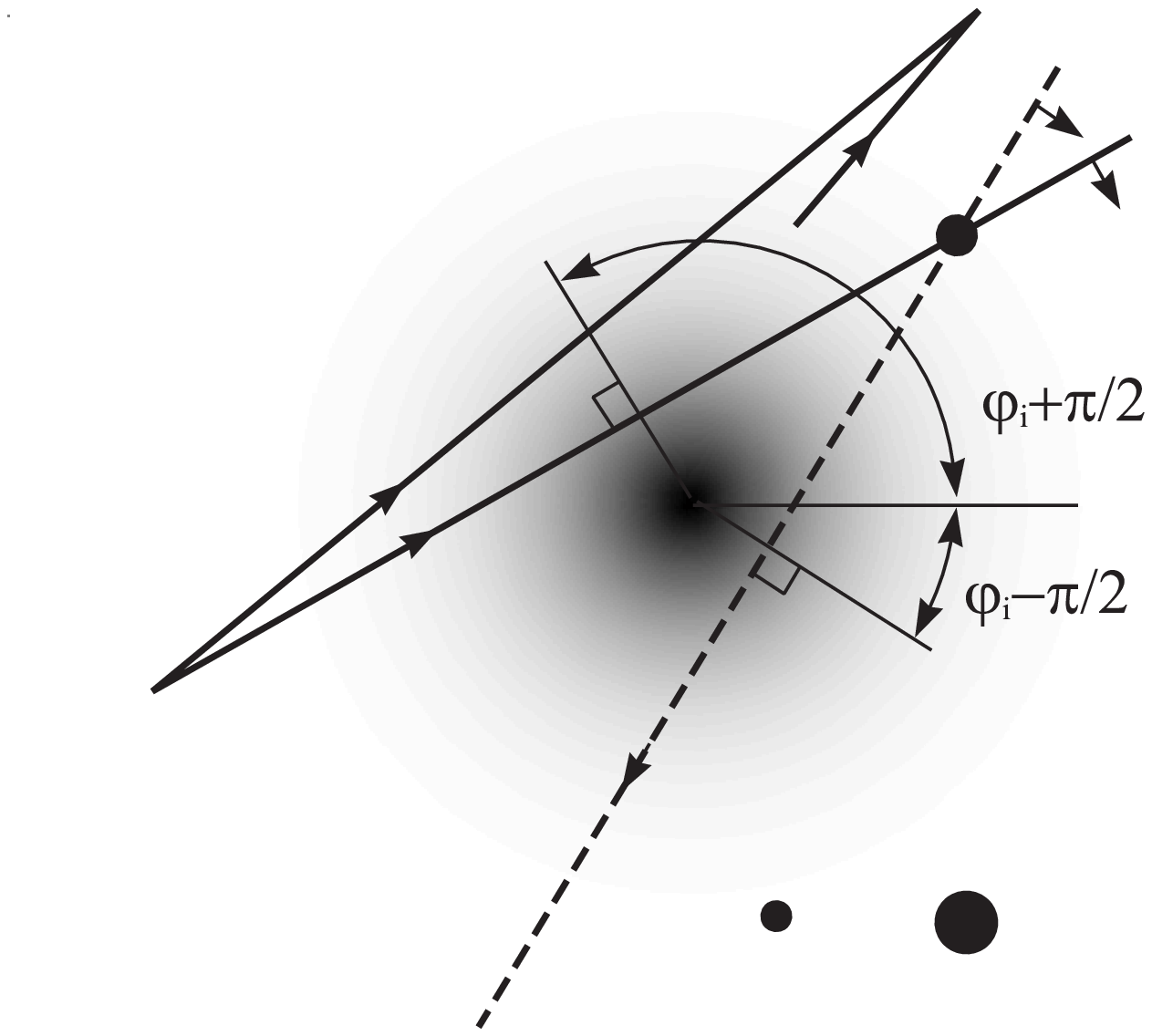,height=1.9in,bbllx=123pt,bblly=340pt,bburx=464pt,bbury=664pt}
}
\vspace{0.1in} 
\setlength{\columnwidth}{3.2in}
\centerline{\caption{
Motion of a quasiparticle within the vortex core. 
The center of the core is shown by the dark region.
The solid line shows a piece of the trajectory of the quasiparticle. 
The arrow indicates the direction of momentum of the quasiparticle, which is unchanged
by the Andreev reflection. Impurity scattering however changes it to the opposite.
The impurities are shown by the black spots. When colliding with an impurity
the quasiparticle scatters into the state, which is almost diametrically 
opposite (dashed line).
\label{fig10}
}}
\vspace{-0.1in}
\end{figure}

The physical picture behind this equation is as follows: the quasiparticle
flies back and forth almost in a straight line, Andreev
reflecting from the walls of the vortex core~\cite{Volovik96}
(see Fig.~\ref{fig05}).
This motion is performed with the characteristic 
frequency $v_{\rm F}/\xi \sim \Delta$, which is 
relatively large. In addition to this the quasiparticle performs a slow 
precession with the frequency $\omega_0 \ll \Delta$, described
by the first term in Eq.~(\ref{ExactIntegralEq}).

The second term in Eq.~(\ref{ExactIntegralEq}) 
describes scattering of the quasiparticles by impurities.
Let us first consider a core with only {\em one} impurity in it.~\cite{Larkin97}
It is typically situated at a distance $a_i \sim \xi$ from the center of the core.
In this case the parameter $k_{\rm F}a_i$ is large. The scattering matrix 
(\ref{ScatteringMatrix}) is then localized at the angles $\phi_i \pm \pi/2$,
with the characteristic spread
$\delta\phi_i \sim 1/\sqrt{k_{\rm F}a_i} \ll 1$.  
This means that the impurity scatters the quasiparticle from 
$\phi = \phi_i+\pi/2$ to $\phi_i-\pi/2$ and vice versa
(see Fig.~\ref{fig10}). Eq.~(\ref{ExactIntegralEq}) can then be integrated
near these points to render the boundary conditions for the wavefunction at them.
To present these boundary conditions it is convenient 
to introduce instead of one function $\Psi(\phi)$ the spinor 
\begin{equation}
\hat{\Psi} \equiv \left(
\begin{array}{l}
{\displaystyle
\Psi(\phi)
} \\ 
{\displaystyle
\Psi(\phi+\pi)
}
\end{array}
\right),
\label{SpinorDef}
\end{equation}  
which is naturally defined for $0 \le \phi < \pi$.
Then the boundary conditions read:
\begin{equation}
\hat{\Psi}\left(\phi_i + \pi/2 + \nu\right)=
M_i\hat{\Psi}\left(\phi_i + \pi/2 - \nu\right),
\label{BC}
\end{equation}
where $\nu \gg \delta \phi_i$. Solving Eqs. (\ref{ExactIntegralEq}) and (\ref{ScatteringMatrix}) 
near impurity we obtain the transfer matrix at the impurity
\begin{equation}
M_i=\frac {1}{1+\left|J\right|^2}
\left(
\begin{array}{ll}
{\displaystyle 1-\left|J\right|^2} & {\displaystyle -2J}\\ 
{\displaystyle 2J^*} & {\displaystyle 1-\left|J\right|^2}\\ 
\end{array}
\right),
\label{DefinitionMi}
\end{equation}
where
$
J={V_i} e^{2ik_{\rm F} a_i-2K\left(a_i\right)}/{2\pi\omega_0 a_i\xi}.
\label{DefinitionJ}
$
%If the angle along which the quasiparticle is precessing is
%considered a ``time'' variable, the boundary conditions (\ref{BC})
%relate the wavefunction of the quasiparticle ``before'' and ``after''
%collision with the impurity.
Noticing that between collisions with impurities
according to Eq.~(\ref{ExactIntegralEq})
$
\hat{\Psi} \propto \exp\left[i\left(E/\omega_0-1/2\right)\phi\right],
\label{FreeEvolution}
$
%As it is easy to notice, the matrix $M_i$ is unitary. 
%Scattering by impurities therefore
%does not change the norm of the wavefunction $\hat{\Psi}^{\dagger}\hat{\Psi}$.
%Neither does the evolution of the wavefunction between impurities 
%(\ref{FreeEvolution})  
%as it only leads to a multiplication by a complex number 
%$\exp\left[i\left(E-\omega_0/2\right)\Delta\phi\right]$,
%where $\Delta\phi$ is the angle traveled. We conclude that
%despite of the one-dimensionality of the problem there cannot be localization
%of the wavefunction in this approximation.
and the boundary conditions for spinor $\hat{\Psi}$ at points
$\phi=0$ and $\pi$ are
$
\hat{\Psi}(\pi-0) = \sigma_x \hat{\Psi}(+0)
\label{BCatZero}
$, we obtain the equation for the excitation energy levels
in the presence of one impurity
\begin{equation}
\cos\left(\pi E/\omega_0\right)= - i{\rm Tr} \left( \sigma_x M_i \right) /2.
\label{CosinusOne}
\end{equation}
This equation coincides with the one derived earlier 
in Ref.~\onlinecite{Larkin97}.

This approach can be generalized to include many impurities. 
In this case the transfer matrices corresponding to different impurities
should be multiplied to provide the transfer matrix from $\phi=0$ to $\pi$.
The resulting equation for the energy levels inside the vortex core is
analogous to (\ref{CosinusOne})
\begin{equation}
\cos\left(\pi E/\omega_0\right)= - i{\rm Tr} \left(\sigma_x \prod_i M_i \right) / 2.
\label{CosinusMany}
\end{equation}
This equation is the central result of this paper.
It is valid in the limit $\lambda_{\rm F} \rightarrow 0$.
More precisely 
$N_i\delta \phi_i \sim N_i(\lambda_{\rm F}/\xi)^{1/2}\ll 1$. Here
$N_i\sim n_i\xi^2$ is the number of impurities in the vortex core and
$n_i$ is the density of impurities.

From Eq.~(\ref{CosinusMany}) it follows that the spectrum of the 
system is periodic with the period $2\omega_0$. It is 
also symmetric with respect to the replacement $E \rightarrow -E$
(particle-hole symmetry).
Let us study the properties of this spectrum in the limit $N_i \gg 1$.
The unitary matrix $M_i$ is a quaternion. Therefore it can be written
in the form $M_i = n_{0i} \sigma_0 + i\bbox{n}_i\bbox{\sigma}$, where
$\sigma_0$ is the unit 2 by 2 matrix, and 
$n_i = \left(n_{0i}, \bbox{n}_i\right)$ is a four-dimensional (4D) 
unit vector. 
The product of the quaternions is also a quaternion. 
Therefore
\begin{equation}
\begin{array}{ll}
{\displaystyle
M \equiv \prod_i M_i \equiv n_0\sigma_0 + i\bbox{n}\bbox{\sigma},} \\ 
{\displaystyle
\cos \left(\pi E/\omega_0\right)=n_x,}
\end{array}
\label{DefinitionOfn}
\end{equation}
where the unit 4-vector $n$ parameterize the matrix $M$.
Every impurity produces a rotation of this vector. 
After the action of many 
impurities the probability to find it at a particular
point on 4D sphere $n^2=1$ spreads uniformly over the 
3D surface of this sphere. Therefore averaging over the position
of impurities can be replaced by averaging over the 4D unit sphere.
Eq.~(\ref{CosinusMany}) results in the following expression
for the average DOS
\begin{equation}
\rho(E)=\int d^4n \delta\left( n^2-1 \right)
\delta\left[ E - {\omega_0}{\rm arccos} \left(n_x\right)/{\pi}\right]
\label{DOS}
\end{equation}  
This results in Eq.~(\ref{IntroDOS}) above.

We now turn to the calculation of vortex viscosity in this model.
The vortex motion produces transitions between
excited states within the vortex core.
These transitions result in the energy dissipation
equal to $\omega_0 W$, where $W$ is
the transition rate between nearest levels.~\cite{Wilkinson92}
Viscosity $\eta$ can be related to the dissipation rate 
per vortex core $Q$ through:
\begin{equation}
\eta = Q/v^2 = \omega_0 W/v^2.
\label{ViscosityDef}
\end{equation}
Here $v$ is the velocity of vortex motion. 
Our task will be to determine $W$. We assume that the velocity
is small, so that the transitions between levels are of 
the Landau-Zener type.~\cite{Feigelman97,Larkin97,Wilkinson92,LL3}
Levels therefore have to come very close to each other.
Due to the particle-hole symmetry of the spectrum 
this implies that energy of the states near Fermi level 
should be very close to zero. From Eq.~(\ref{DefinitionOfn}) 
we see that in this case 
the component $n_x$ of the 4-vector identifying the position 
of matrix $M$ on the 4D sphere should be close to $1$. 
Let us imagine now that the vortex is moving relative to the impurities.
Vector $n$ will then wander over the surface of the 4D sphere
occasionally approaching the pole $n_x=1$, where the Zener 
transitions are possible. Due to rotational symmetry we can
assume that it approaches the pole from the direction 
parallel to $n_0$. Therefore $n_0$ changes sign near the pole
and can be written as $n_0 = v_4t$, where $\bbox{v}_4\equiv \dot{n}$
is the 4-velocity of the point on the sphere. 
Expanding the expression for energy in Eq.~(\ref{DefinitionOfn}) 
we obtain the distance between levels when they are close 
to each other as a function of time:
\begin{equation}
s(t) = 2\omega_0\sqrt{n_y^2+n_z^2+v_4^2t^2}/\pi.
\end{equation}
This allows us to use the Landau-Zener formula\cite{LL3} to calculate
the probability of transitions:
\begin{equation}
W = \int dn_ydn_z v_4 \exp\left(-
\frac{\omega_0\left[ n_y^2+n_z^2 \right]}{v_4} \right)
= \frac{\pi v_4^2}{\omega_0}.
\label{Rate}
\end{equation} 
Here $dn_ydn_z v_4$ is the differential frequency of attempts.
We conclude that the average rate of Landau-Zener transition is 
proportional to the average square of the velocity of vector $n$.
Our next goal will be to evaluate this average.
 
Vector $n$ moves on the 4D sphere because the matrices $M_i$ in 
Eq.~(\ref{DefinitionOfn}) change. The matrices change 
due to the vortex motion relative to the impurities,
according to Eq.~(\ref{DefinitionMi}). 
Noticing that the average velocity $v_4$ is the same all over the sphere
because the density of vector $n$ is uniform there, and that all 4 directions of $v_4$ are equivalent we write
\begin{equation}
\overline{v_4^2} = 4\overline{\dot{n}^2_0} = 4v^2\sum_i
\overline{\left( \partial n_0/\partial x_i \right)^2}.
\end{equation}
The average is taken over the impurity potential.
The latter equation emphasizes that the average of the cross term 
containing the products of two impurities is zero.
To evaluate the derivative with respect to
the coordinate of one impurity $\partial n_0/\partial x_i$
we notice from Eq.~(\ref{DefinitionOfn}) that
\begin{equation}
n_0 = {\rm Tr} \left(M_i R \right)/2.
\end{equation}
Here $R$ is a random 2 by 2 unitary matrix obtained from $M$
by cyclic permutations under the trace. Considering then a small
displacement of the impurity $\delta x_i$ we obtain
\begin{equation}
\delta n_0 = 4k_{\rm F} 
\frac{{\rm Im} \left( J^{*}R_{12}\right)}
{1+|J|^2}\frac{x_i}{a_i}\delta x_i.
\end{equation}
We then average $(\delta n_0 /\delta x_i)^2$ 
over the random matrix $R$ instead of impurities:
\begin{equation}
\overline{v_4^2}=\int_0^{\infty} n_i2\pi rdr 
\frac {8v^2k_{\rm F}^2|J(r)|^2}{\left( 1+|J|^2\right)^2}=
\frac{2\pi v^2k_{\rm F}^2}{\omega_0\tau}
\ln\frac{1}{\theta}.
\end{equation}
Using this result, Eqs.~(\ref{ViscosityDef}), (\ref{Rate}), 
and the connection between viscosity and conductivity
$\sigma_{xx} = |e|c\eta/\pi B$, where $e$ is the charge of electron,
we obtain Eq.~(\ref{Viscosity}) above. 
%Notice that this result 
%significantly exceeds the result of kinetic 
%equation in the $\tau$ approximation (\ref{BardeenStephen}). 

In our consideration we ignored the energy relaxation processes,
assuming the inelastic collision time $\tau_{\varepsilon}$ to be large.
More precisely, we have assumed that  
$v \gg v_{\varepsilon}\sim\lambda_{\rm F}\sqrt{\tau \omega_0}/\tau_{\varepsilon}$.
In the opposite case $v \ll v_{\varepsilon}$, Eq.~(\ref{Viscosity}) 
should be multiplied by a small factor 
$v/v_{\varepsilon}$. At large velocities our approximation 
fails when the transitions to the next nearest neighbor
levels become possible, i.e. $v \gg \lambda_{\rm F}\sqrt{\tau \omega_0^3}$. 
Our approach therefore has an interval of validity if $\omega_0\tau_{\varepsilon} \gg 1$. 
We have also ignored the overheating
of the core, assuming $\tau_{\varepsilon} v^2 \ll \tau \lambda_{\rm
F}^2\Delta^2$, and pinning, assuming that either the observations are made at a frequency
exceeding the pinning frequency~\cite{Koulakov98} or the vortex lattice is rigid. 

The well-known examples of clean layered superconductors are HTSC. 
For them both the vortex DOS\cite{Hess} and viscosity\cite{Ong} are measured. 
Due to the nodes in the superconducting gap and the absence of the Fermi spectrum in the directions 
distant from the nodes, even with no impurities the spectrum of vortex states in HTSC is not clear.
The applicability of our results to this case is not obvious and warrants further study.
We believe however that our conclusion about the existence of the circular random matrix ensemble
is likely to be true for the d-wave superconductors. This is justified by the recent 
study of clean d-wave superconductors, revealing the $2\omega_0$ periodicity.\cite{Volovik97}
We believe therefore that DOS given by Eq.~(\ref{IntroDOS}) may be observed, 
should our regime be accessible experimentally. 

In conclusion, we have studied the DOS and absorption
by the vortex core states in clean layered s-wave superconductors, 
when $\theta^2\Delta\sqrt{\Delta/E_{\rm F}} \gg 1/\tau \gg \omega_0$.
We predict the existence of the circular random matrix ensemble in this regime.
We have found that both DOS and the dissipative component
of conductivity are different from the isotropic 3D case.
This difference is due to the fact that in 2D the quasiparticles
can scatter coherently on the same impurity profile many times.

The authors are grateful to M.~V.~Feigel'man, V.~E.~Kravtsov,
B.~I.~Shklovskii, and M.~A.~Skvortsov for useful discussions.
A.~K. was supported by NSF grant DMR-9616880, A.~L. by NSF grant DMR-9812340.

\vspace{-0.2in}
%-------------------------------------------------------------------------------

\end{multicols}
\end{document}